\documentclass[showpacs,twocolumn,showkeys,preprintnumbers,amsmath,amssymb]{revtex4}

\usepackage{graphicx}
\usepackage{dcolumn}

\begin{document}
\title{Interaction of a two-level atom with squeezed light}
\author{Eyob Alebachew }
\email{yob_a@yahoo.com}
\author{K. Fesseha}
\affiliation{Department of Physics, Addis Ababa University, P. O.
Box 33085, Addis Ababa, Ethiopia}

\date{\today}

\begin{abstract}
We consider a degenerate parametric oscillator whose cavity
contains a two-level atom. Applying the Heisenberg and quantum
Langevin equations, we calculate in the bad-cavity limit the mean
photon number, the quadrature variance, and the power spectrum for
the cavity mode in general and for the signal light and
fluorescent light in particular. We also obtain the normalized
second-order correlation function for the fluorescent light. We
find that the presence of the two-level atom leads to a decrease
in the degree of squeezing of the signal light. It so turns out
that the fluorescent light is in a squeezed state and the power
spectrum consists of a single peak only.
\end{abstract}

\pacs{42.50.Dv, 42.50.Lc, 42.50.Ar}

\keywords{ Photon antibunching; Quadrature variance; Power
spectrum}

\maketitle
\section{Introduction}
A considerable interest has been shown in the analysis of the
effects of squeezed light on the quantum properties of the
fluorescent light emitted by a two-level atom in a cavity. The
power spectrum of the fluorescent light emitted by a two-level
atom interacting with a cavity mode driven by coherent light and
coupled to a squeezed vacuum reservoir has been studied by several
authors~\cite{1,2,3,4,5,6,7}. Some of these studies show that the
width of the incoherent spectrum in the weak driving light limit
decreases as the degree of squeezing increases~\cite{4,7}. On the
other hand, for a strong driving light, the side peaks of the
Mollow spectrum are always broadened while the central peak could
be broadened or narrowed depending on the relative phase between
the strong driving light and the squeezed vacuum~\cite{4,7}.
Moreover, Agarwal \cite{8} has considered coherently driven $N$
two-level atoms passing through a squeezed cavity mode in the
good-cavity limit. He has found modifications of the Mollow
triplet due to the presence of the squeezed light. On the other
hand, Jin and Xiao \cite{9} have considered $N$ two-level atoms
placed inside a parametric oscillator in the good-cavity limit.
They have found that under strong-interaction limit, the presence
of the two-level atoms inside the parametric oscillator increases
the amount of intracavity squeezing from its maximum value of
$50\%$ to a maximum value of $75\%$. In addition, Clemens {\it et
al.}~\cite{10} have investigated the power spectrum of the light
emitted by a two-level atom inside a parametric oscillator in the
week driving light limit. They have found that the incoherent
spectrum consists of a vacuum-Rabi doublet with holes in each side
band.

In this paper we consider a degenerate parametric oscillator
operating below threshold and whose cavity contains a two-level
atom. The interaction of the signal light, produced by the
parametric amplifier, with the two-level atom leads to the
generation of fluorescent light. Thus the cavity mode in this case
consists of the signal light and the fluorescent light emitted by
the two-level atom. In this paper we analyze the quantum
statistical properties of the fluorescent and the signal light
applying the Heisenberg and quantum Langevin equations in the
bad-cavity limit. This system can also be studied using the master
equation in the bad-cavity limit. Employing the bad-cavity limit,
one usually obtains the master equation for the atomic density
operator. Hence it will not be possible in this approach to study
the properties of the cavity mode. The method used in this paper
enables us to study not only the properties of the fluorescent
light emitted by the two-level atom but also the properties of the
cavity mode.

We derive the equations of evolution for the expectation values of
atomic and cavity mode operators using the Heisenberg and quantum
Langevin equations in the bad-cavity limit. Applying the resulting
equations, we calculate the mean photon number, the quadrature
variance, and the power spectrum for the cavity mode, for the
signal light, and for the fluorescent light. We also determine the
second order correlation function for the fluorescent light.

\section{Equations of Evolution of atomic expectation values}
We consider a single two-level atom inside a parametric oscillator
coupled to a vacuum reservoir. We represent the upper and lower
levels of the atom by $|a\rangle$ and $|b\rangle$ and we assume
the atom to be at resonance with the cavity mode (see Fig. 1). In
a degenerate parametric oscillator, a pump photon of frequency
$2\omega$ is down converted into a pair of highly correlated
signal photons each of frequency $\omega$. It so turns out that
the signal light is in a squeezed state. Contrary to the work of
Clemens {\it{ et al}}. \cite{10} where they considered weak
squeezed light (two photons in the cavity at a time), we have not
imposed any restriction on the number of signal photons in the
cavity. With the pump mode treated classically, the parametric
interaction can be described by the Hamiltonian ~\cite{11}
\begin{figure}
\includegraphics [height=3.5cm,angle=0]{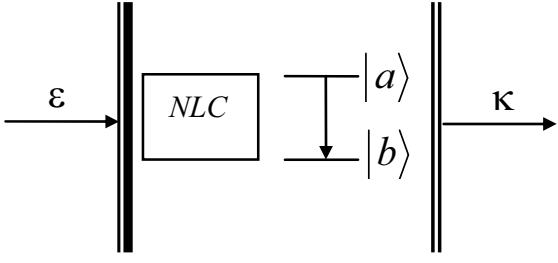}
\caption{A single two-level atom inside a parametric oscillator.}
\end{figure}
\begin{equation}\label{1}
\hat H_{1}=\frac{i\varepsilon}{ 2}(\hat a^{\dagger 2}-\hat a^2),
\end{equation}
in which $\varepsilon$, assumed to be real and constant, is
proportional to the amplitude of the pump mode and $\hat a$ is the
annihilation operator for the cavity mode.  In addition, the
interaction of the cavity mode with the two-level atom is
describable by the Hamiltonian
\begin{equation}\label{2}
\hat H_{2}=ig(\hat\sigma_{+}\hat a-\hat a^{\dagger}\sigma_{-}),
\end{equation}
where $g$ is the atom-cavity mode coupling constant and
$\hat\sigma_{\pm}$ are atomic operators satisfying the commutation
relations $[\hat\sigma_{+},\hat\sigma_{-}]=\hat\sigma_{z}$  and
$[\hat\sigma_{\pm},\hat\sigma_{z}]=\mp 2\hat\sigma_{\pm}$. Thus
the Hamiltonian describing the parametric interaction and the
interaction of the cavity mode with the two-level has the form
\begin{equation}\label{3}
H={i\varepsilon\over 2}(\hat a^{\dagger 2}-\hat
a^2)+ig(\hat\sigma_{+}\hat a-\hat a^{\dagger}\sigma_{-}).
\end{equation}
Applying the Heisenberg equation, one can readily establish that
the time evolution of the atomic operators are of the form
\begin{equation}\label{4}
\frac{d}{dt}\hat\sigma_{-}=-g\hat\sigma_{z}\hat a,
\end{equation}
\begin{equation}\label{5}
\frac{d}{dt}\hat\sigma_{z}=2g\hat
a^{\dagger}\hat\sigma_{-}+2g\hat\sigma_{+}\hat a.
\end{equation}
On the other hand, the quantum Langevin equation for the cavity
mode operator $\hat a$ is expressible as
\begin{subequations}
\begin{equation}\label{6a}
\frac{d}{dt}\hat a=-i[\hat a,\hat H]-\frac{\kappa}{2}\hat a+\hat
F,
\end{equation}
so that on account of Eq. \eqref{3}, there follows
\begin{equation}\label{6b}
\frac{d}{dt}\hat a=-\frac{\kappa}{2}\hat a+\varepsilon \hat
a^{\dagger}-g\hat\sigma_{-}+\hat F,
\end{equation}
\end{subequations}
where $\kappa$ is the cavity damping constant and $\hat F$ is a
noise operator associated with the vacuum reservoir and having the
following correlation properties:
\begin{subequations}\label{7}
\begin{equation}\label{7a}
\langle \hat F(t)\rangle=0,
\end{equation}
\begin{equation}\label{7b}
\langle \hat F^{\dagger}(t)\hat F(t^{\prime})\rangle=0,
\end{equation}
\begin{equation}\label{7c}
\langle \hat F(t)\hat
F^{\dagger}(t^{\prime})\rangle=\kappa\delta(t-t^{\prime}),
\end{equation}
\begin{equation}\label{7d}
\langle \hat F^{\dagger}(t)\hat
F^{\dagger}(t^{\prime})\rangle=\langle \hat F(t)\hat
F(t^{\prime})\rangle=0.
\end{equation}

Since Eqs. \eqref{4}, \eqref{5}, and \eqref{6b} are nonlinear and
coupled differential equations, it is not possible to obtain exact
solutions. We then seek to obtain the solutions of these equations
applying the bad-cavity limit. In the bad-cavity limit, the cavity
damping constant is much greater than the cavity atomic decay
rate. In this limit, the cavity mode variables decay faster than
the atomic variables. We can then set the time derivatives of the
cavity mode variables equal to zero while keeping the zero-order
atomic and cavity mode variables at time t. In view of this, we
obtain from Eq. \eqref{6b} that
\end{subequations}
\begin{align}\label{8}
\hat a(t)=&-\frac{2\kappa
g}{\kappa^2-4\varepsilon^2}\hat\sigma_{-}(t)-\frac{4
g\varepsilon}{\kappa^2-4\varepsilon^2}\hat\sigma_{+}(t)\notag\\
&+\frac{4}{\kappa^2-4\varepsilon^2}(\frac{\kappa}{2}\hat
F(t)+\varepsilon\hat F^{\dagger}(t)).
\end{align}
This result will be used to calculate the expectation values of
the products of a cavity mode operator and an atomic operator.
Then introduction of Eq. \eqref{8} into \eqref{4} and \eqref{5}
leads to
\begin{align}\label{9}
\frac{d}{dt}\hat\sigma_{-}&=-\frac{2g^2/\kappa}{1-4\varepsilon^2/\kappa^2}\hat\sigma_{-}
+\frac{4g^2\varepsilon/\kappa^2}{1-4\varepsilon^2/\kappa^2}\hat\sigma_{+}\notag\\
&-\frac{4g}{\kappa^2-4\varepsilon^2}\big[\frac{\kappa}{2}\hat\sigma_{z}\hat
F+\varepsilon\hat\sigma_{z}\hat F^{\dagger}\big],
\end{align}
\begin{align}\label{10}
\frac{d}{dt}\hat\sigma_{z}&=-\frac{8g^2/\kappa}
{1-4\varepsilon^2/\kappa^2}\hat\sigma_{+}\hat\sigma_{-}\notag\\
&+\frac{8g}{\kappa^2-4\varepsilon^2}\big[\frac{\kappa}{2}(\hat
F^{\dagger}\hat\sigma_{-}+\hat\sigma_{+}\hat F)+\varepsilon(\hat
F\hat\sigma_{-}+\hat\sigma_{+}\hat F^{\dagger})\big],
\end{align}
or
\begin{align}\label{11}
\frac{d}{dt}\langle\hat\sigma_{-}\rangle &=-\frac{2g^2/\kappa}
{1-4\varepsilon^2/\kappa^2}\langle\hat\sigma_{-}\rangle
+\frac{4g^2\varepsilon/\kappa^2}
{1-4\varepsilon^2/\kappa^2}\langle\hat\sigma_{+}\rangle\notag\\
&-\frac{4g}{\kappa^2-4\varepsilon^2}\big[\frac{\kappa}{2}\langle\hat\sigma_{z}\hat
F\rangle+\varepsilon\langle\hat\sigma_{z}\hat
F^{\dagger}\rangle\big],
\end{align}
\begin{align}\label{12}
&\frac{d}{dt}\langle\hat\sigma_{z}\rangle=-\frac{8g^2/\kappa}
{1-4\varepsilon^2/\kappa^2}\langle\hat\sigma_{+}\hat\sigma_{-}\rangle\notag\\
&+\frac{8g}{\kappa^2-4\varepsilon^2}\big[\frac{\kappa}{2}(\langle\hat
F^{\dagger}\hat\sigma_{-}\rangle+\langle\hat\sigma_{+}\hat
F\rangle)+\varepsilon(\langle\hat
F\hat\sigma_{-}\rangle+\langle\hat\sigma_{+}\hat
F^{\dagger}\rangle)\big].
\end{align}
We note that Eq. \eqref{9} has a well-behaved solution provided
that $\eta=(4g^2/\kappa)/(1-4\varepsilon^2/\kappa^2)$ is positive.
This will be the case if $\varepsilon/\kappa<1/2$.

We next proceed to find the expectation values of the products
involving a noise operator and an atomic operator that appear in
Eqs. \eqref{11} and \eqref{12}. To this end, the formal solution
of Eq. \eqref{9} can be written as
\begin{align}\label{13}
\hat\sigma_{-}(t)&=\hat\sigma_{-}(0)e^{-\eta
t/2}\notag\\
&+\int_{0}^{t}e^{-\eta(t-t^{\prime})/2}\big[
\eta(\varepsilon/\kappa)\hat\sigma_{+}(t^{\prime})
-\frac{4g}{\kappa^2-4\varepsilon^2}\notag\\
&\times\big(\frac{\kappa}{2}\hat\sigma_{z}(t^{\prime})\hat
F(t^{\prime})+\varepsilon\hat\sigma_{z}(t^{\prime})\hat
F^{\dagger}(t^{\prime})\big)\big]dt^{\prime},
\end{align}
so that multiplying this equation on the left by $\hat F(t)$ and
taking the expectation value of the resulting expression, we
obtain
\begin{align}\label{14}
&\langle\hat F(t)\hat\sigma_{-}(t)\rangle =\langle\hat
F(t)\hat\sigma_{-}(0)\rangle e^{-\eta
t/2}\notag\\&+\int_{0}^{t}e^{-\eta(t-t^{\prime})/2}\big[
\eta(\varepsilon/\kappa)\langle\hat
F(t)\hat\sigma_{+}(t^{\prime})\rangle
-\frac{4g}{\kappa^2-4\varepsilon^2}\notag\\
&\times\big(\frac{\kappa}{2}\langle\hat
F(t)\hat\sigma_{z}(t^{\prime})\hat
F(t^{\prime})\rangle+\varepsilon\langle\hat
F(t)\hat\sigma_{z}(t^{\prime})\hat
F^{\dagger}(t^{\prime})\rangle\big)\big]dt^{\prime}.
\end{align}
It is not possible to evaluate the integral that appears in Eq.
\eqref{14} as the explicit form of $\hat\sigma_{z}(t^{\prime})$ is
unknown yet. In order to proceed further, we need to adopt a
certain approximation scheme. To this end, ignoring the
noncommutativity of the atomic and noise operators, we see that
$\langle\hat F(t)\hat\sigma_{z}(t^{\prime})\hat
F(t^{\prime})\rangle=\langle\hat\sigma_{z}(t^{\prime})\hat
F(t)\hat F(t^{\prime})\rangle$. Then upon neglecting the
correlation between $\hat\sigma_{z}(t^{\prime})$ and $\hat
F(t)\hat F(t^{\prime})$, assumed to be considerably small, one can
write the approximately valid relation~\cite{12} $\langle\hat
F(t)\hat\sigma_{z}(t^{\prime})\hat
F(t^{\prime})\rangle=\langle\hat\sigma_{z}(t^{\prime})\hat
F(t)\hat
F(t^{\prime})\rangle=\langle\hat\sigma_{z}(t^{\prime})\rangle\langle\hat
F(t)\hat F(t^{\prime})\rangle$. Following a similar line of
reasoning, one can also write the approximately valid relation
$\langle\hat F(t)\hat\sigma_{z}(t^{\prime})\hat
F^{\dagger}(t^{\prime})\rangle=\langle\hat\sigma_{z}(t^{\prime})\hat
F(t)\hat
F^{\dagger}(t^{\prime})\rangle=\langle\hat\sigma_{z}(t^{\prime})\rangle\langle\hat
F(t)\hat F^{\dagger}(t^{\prime})\rangle$.  Now using these
approximations and taking into account the fact that a noise
operator $\hat F$ at time t does not affect the atomic variables
at earlier times, Eq. \eqref{14} can be put in the form
\begin{align}\label{15}
\langle\hat
F(t)\hat\sigma_{-}(t)\rangle&=-\frac{4g}{\kappa^2-4\varepsilon^2}
\int_{0}^{t}e^{-\eta(t-t^{\prime})/2}\langle\hat\sigma_{z}(t)^{\prime}\rangle\notag\\
&\times\big(\frac{\kappa}{2} \langle\hat F(t)\hat
F(t^{\prime})\rangle+ \varepsilon\langle\hat F(t)\hat
F^{\dagger}(t^{\prime})\rangle\big)dt^{\prime}.
\end{align}
Therefore using Eqs. \eqref{7c} and \eqref{7d} and performing the
integration, we find
\begin{equation}\label{16}
\langle\hat
F(t)\hat\sigma_{-}(t)\rangle=-\frac{(2g\varepsilon/\kappa)\langle\hat\sigma_{z}(t)\rangle}
{1-4\varepsilon^2/\kappa^2}.
\end{equation}
We immediately notice that
\begin{equation}\label{17}
\langle\hat\sigma_{+}(t)\hat
F^{\dagger}(t)\rangle=-\frac{(2g\varepsilon/\kappa)\langle\hat\sigma_{z}(t)\rangle}
{1-4\varepsilon^2/\kappa^2}.
\end{equation}
It can also be readily established in a similar manner that
\begin{subequations}\label{18}
\begin{equation}\label{18a}
\langle\hat\sigma_{+}(t)\hat F(t)\rangle=\langle\hat
F^{\dagger}(t)\hat\sigma_{-}(t)\rangle=0,
\end{equation}
\begin{equation}\label{18b}
\langle\hat\sigma_{-}(t)\hat F(t)\rangle=\langle\hat
F^{\dagger}(t)\sigma_{+}(t)\rangle=0,
\end{equation}
\end{subequations}
\begin{equation}\label{19}
\langle\hat\sigma_{-}(t)\hat F^{\dagger}(t)\rangle=\langle\hat
F(t)\hat\sigma_{+}(t)\rangle=-\frac{g\langle\hat\sigma_{z}(t)\rangle}{1-4\varepsilon^2/\kappa^2},
\end{equation}
\begin{equation}\label{20}
\langle\hat\sigma_{z}(t)\hat F(t)\rangle=0,
\end{equation}
\begin{equation}\label{21}
\langle\hat\sigma_{z}(t)\hat
F^{\dagger}(t)\rangle=\frac{4g/\kappa}{1-4\varepsilon^2/\kappa^2}(\frac{\kappa}{2}
\langle\hat\sigma_{+}\rangle+\varepsilon\langle\hat\sigma_{-}\rangle).
\end{equation}
With the aid of Eqs. \eqref{16}-\eqref{18}, \eqref{20},
\eqref{21}, and employing the relation
$\hat\sigma_{+}\hat\sigma_{-}=(\hat\sigma_{z}+1)/2$, Eqs.
\eqref{11} and \eqref{12} can be written as
\begin{align}\label{22}
\frac{d}{dt}\langle\hat\sigma_{-}\rangle=&-\frac{\Gamma}{2}
\langle\hat\sigma_{-}\rangle
-\frac{\varepsilon}{\kappa}\Gamma\langle\hat\sigma_{+}\rangle,
\end{align}
\begin{equation}\label{23}
\frac{d}{dt}\langle\hat\sigma_{z}\rangle=-\Gamma
\langle\hat\sigma_{z}\rangle -\eta,
\end{equation}
where $\Gamma=(4
g^2/\kappa)(1+4\varepsilon^2/\kappa^2)/(1-4\varepsilon^2/\kappa^2)^2$
is the cavity atomic decay rate. In view of the fact that
$\langle\hat\sigma_{-}(t)\rangle^{*}=\langle\hat\sigma_{+}(t)\rangle$
and
$\langle\hat\sigma_{z}(t)\rangle^{*}=\langle\hat\sigma_{z}(t)\rangle$
one can write
\begin{align}\label{24}
\frac{d}{dt}\langle\hat\sigma_{+}\rangle=&-\frac{\Gamma}{2}
\langle\hat\sigma_{+}\rangle
-\frac{\varepsilon}{\kappa}\Gamma\langle\hat\sigma_{-}\rangle.
\end{align}
In the absence of the parametric amplifier the cavity atomic decay
rate is $\gamma_{c}=4g^2/\kappa$. Thus we can express $\Gamma$ as
$\Gamma=\gamma_{c}(1+4\varepsilon^2/\kappa^2)/(1-4\varepsilon^2/\kappa^2)^2$.
It can be easily seen that the presence of the parametric
amplifier enhances the cavity atomic decay rate.
\section{Power spectrum and photon antibunching of the fluorescent light}
The power spectrum of the fluorescent light can be expressed as
\cite{12}
\begin{equation}\label{25}
S(\omega)=2Re\int_{0}^{\infty}\langle\hat\sigma_{+}(t)\hat\sigma_{-}(t+\tau)\rangle_{ss}e^{i\omega\tau}d\tau.
\end{equation}
Introducing new variables defined by
$z_{\pm}=\langle\hat\sigma_{-}\rangle\pm\langle\hat\sigma_{+}\rangle$
and applying Eqs. \eqref{22} and \eqref{24}, we get
\begin{equation}\label{26}
\frac{d}{dt}z_{\pm}=-\lambda_{\pm}z_{\pm},
\end{equation}
where
$\lambda_{\pm}=\Gamma(\frac{1}{2}\pm\frac{\varepsilon}{\kappa})$.
The solution of this equation can be written in the form
\begin{equation}\label{27}
z_{\pm}(t+\tau)=z_{\pm}(t)e^{-\lambda_{\pm}\tau}.
\end{equation}
It then follows that
\begin{align}\label{28}
\langle\hat\sigma_{-}(t+\tau)\rangle
&=\frac{1}{2}\langle\hat\sigma_{-}(t)\rangle
(e^{-\lambda_{+}\tau}+e^{-\lambda_{-}\tau})\notag\\
&+\frac{1}{2}\langle\hat\sigma_{+}(t)\rangle
(e^{-\lambda_{+}\tau}-e^{-\lambda_{-}\tau}).
\end{align}
Now applying the quantum regression theorem, we have
\begin{equation}\label{29}
\langle\hat\sigma_{+}(t)\hat\sigma_{-}(t+\tau)\rangle_{ss}
=\frac{1}{2}\langle\hat\sigma_{+}(t)\hat\sigma_{-}(t)\rangle_{ss}
(e^{-\lambda_{+}\tau}+e^{-\lambda_{-}\tau}).
\end{equation}
The steady state solution of Eq. \eqref{23} is
\begin{equation}\label{30}
\langle\hat\sigma_{z}(t)\rangle_{ss}=\frac{-\eta}{\Gamma},
\end{equation}
from which follows
\begin{equation}\label{31}
(\langle\hat\sigma_{z}(t)\rangle_{ss}+1)/2=\frac{\Gamma-\eta}{2\Gamma}.
\end{equation}
In view of the relation
$\langle\hat\sigma_{+}(t)\hat\sigma_{-}(t)\rangle_{ss}
=(\langle\hat\sigma_{z}(t)\rangle_{ss}+1)/2$ and Eq. \eqref{31},
we see that
\begin{equation}\label{32}
\langle\hat\sigma_{+}(t)\hat\sigma_{-}(t)\rangle_{ss}
=\frac{\Gamma-\eta}{2\Gamma}
\end{equation}
 Now upon substituting \eqref{32} into \eqref{29}, we
obtain
\begin{align}\label{33}
\langle\hat\sigma_{+}(t)\hat\sigma_{-}(t+\tau)\rangle_{ss}
=\frac{\Gamma-\eta}{4\Gamma}
(e^{-\lambda_{+}\tau}+e^{-\lambda_{-}\tau}).
\end{align}
On account of this result the power spectrum takes the form
\begin{equation}\label{34}
S(\omega)=\frac{\Gamma-\eta}{2\Gamma}Re
\int_{0}^{\infty}(e^{-(\lambda_{+}-i\omega)\tau}+e^{-(\lambda_{-}-i\omega)\tau})d\tau.
\end{equation}
\begin{figure}
\includegraphics [height=6cm,angle=0]{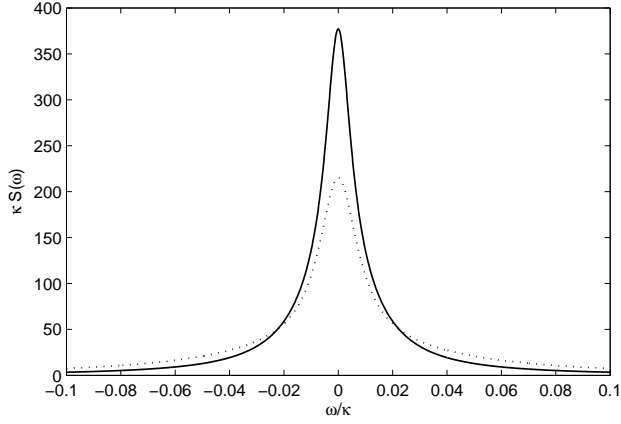}
\caption{Plots of the power spectrum of the fluorescent light [Eq.
\eqref{35}] versus $\omega/\kappa$ for $\gamma_{c}/\kappa=0.01$,
for $\varepsilon/\kappa=0.25$ (solid curve) and for
$\varepsilon/\kappa=0.35$ (dotted curve).}
\end{figure}
Hence the normalized power spectrum is found to be
\begin{equation}\label{35}
S(\omega)=\frac{\Gamma(\frac{1}{2}+\frac{\varepsilon}{\kappa})/2\pi}
{\Gamma^2(\frac{1}{2}+\frac{\varepsilon}{\kappa})^2
+\omega^2}+\frac{\Gamma(\frac{1}{2}-\frac{\varepsilon}{\kappa})/2\pi}
{\Gamma^2(\frac{1}{2}-\frac{\varepsilon}{\kappa})^2+\omega^2}.
\end{equation}
Expression \eqref{35} indicates that the power spectrum of the
fluorescent light is the sum of two Lorentzians centered at zero
frequency and having half widths of
$\Gamma(\frac{1}{2}+\frac{\varepsilon}{\kappa})$ and
$\Gamma(\frac{1}{2}-\frac{\varepsilon}{\kappa})$. Fig. 2 shows
that the power spectrum of the fluorescent light is a single peak
centered at $\omega=0$. We have found that the half width of the
power spectrum increases from $0.0070$ to $0.0101$ as
$\varepsilon/\kappa$ increases from $0.25$ to $0.35$. Contrary to
the power spectrum of the fluorescent light from a two-level atom
driven by a strong coherent light~\cite{4,7}, the power spectrum
in this case turns out to be a single peak.

The second order correlation function can be expressed in terms of
the atomic operators as
\begin{equation}\label{36}
g^{(2)}(\tau)=\frac{\langle\hat\sigma_{+}(t)\hat\sigma_{+}
(t+\tau)\hat\sigma_{-}(t+\tau)\hat\sigma_{-}(t)\rangle}
{\langle\hat\sigma_{+}(t)\hat\sigma_{-}(t)\rangle^2}.
\end{equation}
We recall that
\begin{equation}\label{37}
\langle\hat\sigma_{+}(t+\tau)\hat\sigma_{-}(t+\tau)\rangle=
(\langle\hat\sigma_{z}(t+\tau)\rangle+1)/2.
\end{equation}
Furthermore, the formal solution of Eq. \eqref{23} can be written
as
\begin{equation}\label{38}
\langle\hat\sigma_{z}(t+\tau)\rangle=\langle\hat\sigma_{z}(t)\rangle
e^{-\Gamma\tau}-\frac{\eta}{\Gamma}(1-e^{-\Gamma\tau}),
\end{equation}
from which follows
\begin{align}\label{39}
(\langle\hat\sigma_{z}(t+\tau)\rangle+1)/2&=\frac{1}{2}(\langle\hat\sigma_{z}(t)\rangle+1)
e^{-\Gamma\tau}\notag\\
&+\frac{\Gamma-\eta}{2\Gamma}(1-e^{-\Gamma\tau}).
\end{align}
In view of Eq. \eqref{37}, we see that
\begin{equation}\label{40}
\langle\hat\sigma_{+}(t+\tau)\hat\sigma_{-}(t+\tau)\rangle=
\langle\hat\sigma_{+}(t)\hat\sigma_{-}(t)\rangle
e^{-\Gamma\tau}+\frac{\Gamma-\eta}{2\Gamma}(1-e^{-\Gamma\tau}).
\end{equation}
On applying the quantum regression theorem, the second-order
correlation function can be written as
\begin{equation}\label{41}
g^{(2)}(\tau)= \frac{(\Gamma-\eta)/2\Gamma}
{\langle\hat\sigma_{+}(t)\hat\sigma_{-}(t)\rangle}(1-e^{-\Gamma\tau}).
\end{equation}
Thus in view \eqref{32}, the steady-state second order correlation
function becomes
\begin{equation}\label{42}
g^{(2)}(\tau)= 1-e^{-\Gamma\tau}.
\end{equation}
We observe that $g^{(2)}(0)=0$ and for $\tau > 0$,
$g^{(2)}(\tau)>0$. Therefore we see that for $\tau>0$,
$g^{(2)}(\tau)>g^{(2)}(0)$. The fluorescent light thus exhibits
the phenomenon of photon antibunching, as is always the case. This
is due to the fact that a two-level atom cannot emit two or more
photons simultaneously. After each emission the atom returns to
the lower level and it must absorb a photon before another
emission can take place. Fig. 3 indicates that for relatively
small values of $\tau$ the second-order correlation function is
less than unity which reflects the nonclassical feature of
antibunching. We also observe that as $\varepsilon/\kappa$
increases $g^{(2)}(\tau)$ approaches unity at a faster rate.
\begin{figure}
\includegraphics [height=6cm,angle=0]{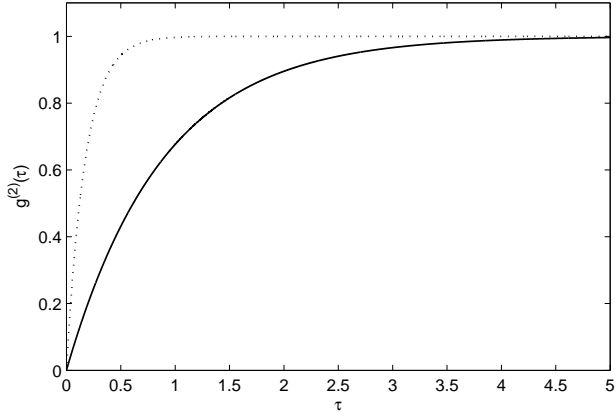}
\caption{Plots of the second order correlation function [Eq.
\eqref{42}] versus $\tau$ for $\gamma_{c}/\kappa=0.01$, for
$\varepsilon/\kappa=0.10$ (solid curve), for
$\varepsilon/\kappa=0.35$ (dotted curve).}
\end{figure}

It is also interesting to consider the dynamics of the two-level
atom. Thus upon replacing $\tau$ by $t$ and $t$ by $0$ in Eq.
\eqref{39} and using the relation
$\langle\hat\sigma_{+}(t)\hat\sigma_{-}(t)\rangle=(\langle\hat\sigma_{z}(t)\rangle+1)/2$,
the probability for the two-level atom to be in the upper level is
found to be
\begin{equation}\label{43}
\rho_{aa}(t)=\rho_{aa}(0)e^{-\Gamma
t}+\frac{4\varepsilon^2/\kappa^2}
{1+4\varepsilon^2/\kappa^2}(1-e^{-\Gamma t}).
\end{equation}
If the atom is initially in the upper level, then
$\rho_{aa}(0)=1$. Hence Eq. \eqref{43} takes for this case the
form
\begin{equation}\label{44}
\rho_{aa}(t)=\frac{e^{-\Gamma t}}{1+4\varepsilon^2/\kappa^2}
+\frac{4\varepsilon^2/\kappa^2}{1+4\varepsilon^2/\kappa^2}
\end{equation}
and at steady state, we have
\begin{equation}\label{45}
\rho_{aa}=\frac{4\varepsilon^2/\kappa^2}{1+4\varepsilon^2/\kappa^2}.
\end{equation}
\begin{figure}
\includegraphics [height=6cm,angle=0]{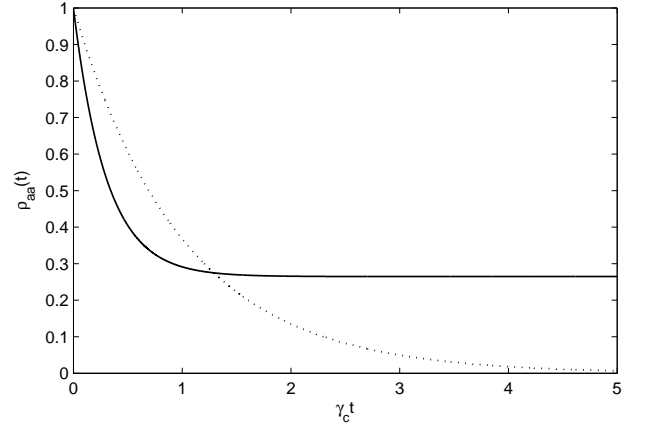}
\caption{Plots of [Eq. \eqref{44}] versus $\gamma_{c}t$ in the
presence of the parametric amplifier with $\varepsilon/\kappa=0.3$
(solid curve) and in the absence of the parametric amplifier, i.e,
for $\varepsilon=0$ (dotted curve).}
\end{figure}
We see from Fig. 4 that the probability for the atom to be in the
upper level decays exponentially in the absence of the parametric
amplifier and approaches to zero at steady state. However, in the
presence of the parametric amplifier the steady state probability
for the atom to be in the upper level is different from zero. This
is because there are photons in the cavity that can be absorbed by
the atom.
\section{Quadrature variance}
In this section we calculate the mean photon number and the
quadrature variance for the cavity mode. Moreover, we determine
the mean photon number and the quadrature variance for the signal
light and for the fluorescent light. The variance of the
quadrature operators defined by ~\cite{11}
\begin{equation}\label{46}
\hat a_{+}=\hat a^{\dagger}+\hat a
\end{equation}
and
\begin{equation}\label{47}
\hat a_{-}=i(\hat a^{\dagger}-\hat a),
\end{equation}
can be expressed as
\begin{equation}\label{48}
\Delta\hat a_{\pm}^2=1\pm(\langle\hat a^{\dagger
^2}\rangle+\langle\hat a^2\rangle\pm 2\langle\hat a^{\dagger}\hat
a\rangle)\mp(\langle\hat a^{\dagger}\rangle\pm\langle\hat
a\rangle)^2.
\end{equation}
On account Eqs. \eqref{7a}, \eqref{8}, and \eqref{28}, we easily
see that
\begin{equation}\label{49}
\langle\hat a^{\dagger}\rangle_{ss}=\langle\hat a\rangle_{ss}=0.
\end{equation}
Thus the quadrature variance takes at steady state the form
\begin{equation}\label{50}
\Delta \hat a_{\pm}^2=1+ 2\langle\hat a^{\dagger}\hat
a\rangle_{ss}\pm(\langle\hat a^{\dagger
^2}\rangle_{ss}+\langle\hat a^2\rangle_{ss}).
\end{equation}

We now proceed to calculate the steady state expectation values of
the second-order cavity mode variables. Employing Eq. \eqref{6b},
one can readily obtain
\begin{align}\label{51}
\frac{d}{dt}\langle\hat a^2\rangle &=-\kappa\langle\hat a^2\rangle
+ 2\varepsilon\langle\hat a^{\dagger}\hat
a\rangle+\varepsilon-g(\langle\hat a
\hat\sigma_{-}\rangle+\langle\hat\sigma_{-}\hat a
\rangle)\notag\\
&+\langle\hat a \hat F\rangle+\langle\hat F\hat a \rangle,
\end{align}
\begin{align}\label{52}
\frac{d}{dt}\langle\hat a^{\dagger}\hat a\rangle
&=-\kappa\langle\hat a^{\dagger}\hat
a\rangle+\varepsilon(\langle\hat a^2\rangle + \langle\hat
a^{\dagger 2}\rangle)\notag\\
& -g(\langle\hat a^{\dagger}
\hat\sigma_{-}\rangle+\langle\hat\sigma_{+}\hat a \rangle)
+\langle\hat a^{\dagger} \hat F\rangle+\langle \hat
F^{\dagger}\hat a\rangle.
\end{align}
The formal solution of Eq. \eqref{6b} can be expressed as
\begin{equation}\label{53}
\hat a(t)=\hat a(0)e^{-\kappa
t/2}+\int_{0}^{t}e^{-\kappa(t-t^{\prime})/2}[\varepsilon\hat
a^{\dagger}(t^{\prime})-g\hat\sigma_{-}(t^{\prime})+\hat
F(t^{\prime})]dt^{\prime},
\end{equation}
so that multiplying on the right by $\hat F(t)$ and taking the
expectation value, we get
\begin{align}\label{54}
\langle\hat a(t)\hat F(t)\rangle &=\langle\hat a(0)\hat
F(t)\rangle e^{-\kappa
t/2}+\int_{0}^{t}e^{-\kappa(t-t^{\prime})/2}[\varepsilon\langle\hat
a^{\dagger}(t^{\prime})\hat
F(t)\rangle\notag\\
&-g\langle\hat\sigma_{-}(t^{\prime})\hat F(t)\rangle+\langle\hat
F(t^{\prime})\hat F(t)\rangle]dt^{\prime}.
\end{align}
On account of Eq. \eqref{7d} and the fact that the noise operator
at time $t$ does not affect the system variables at earlier times,
Eq. \eqref{54} reduces to
\begin{align}\label{55}
\langle\hat a(t)\hat F(t)\rangle =0.
\end{align}
It can also be established in a similar manner that
\begin{align}\label{56}
\langle\hat F(t)\hat a(t)\rangle =0.
\end{align}
Furthermore, applying Eq. \eqref{8} along with Eqs. \eqref{16},
\eqref{18}, and \eqref{19}, one easily obtains
\begin{align}\label{57}
\langle\hat a\hat\sigma_{-}\rangle=
-\frac{4g\varepsilon/\kappa^2}{1-4\varepsilon^2/\kappa^2}\langle\hat\sigma_{+}\hat\sigma_{-}\rangle
-\frac{4g\varepsilon/\kappa^2}{(1-4\varepsilon^2/\kappa^2)^2}\langle\hat\sigma_{z}\rangle
\end{align} and
\begin{align}\label{58}
\langle\hat\sigma_{-}\hat a\rangle=
-\frac{4g\varepsilon/\kappa^2}{1-4\varepsilon^2/\kappa^2}\langle\hat\sigma_{-}\hat\sigma_{+}\rangle
-\frac{4g\varepsilon/\kappa^2}{(1-4\varepsilon^2/\kappa^2)^2}\langle\hat\sigma_{z}\rangle.
\end{align}
Upon substituting Eqs. \eqref{55}-\eqref{58} into \eqref{51}, we
find
\begin{align}\label{59}
\frac{d}{dt}\langle\hat a^2\rangle &=-\kappa\langle\hat a^2\rangle
+ 2\varepsilon\langle\hat a^{\dagger}\hat
a\rangle+\varepsilon+\frac{\gamma_{c}\varepsilon/\kappa}{1-4\varepsilon^2/\kappa^2}\notag\\
&+\frac{2\gamma_{c}\varepsilon/\kappa}{(1-4\varepsilon^2/\kappa^2)^2}\langle\hat\sigma_{z}\rangle.
\end{align}
Following a similar procedure, one can put Eq. \eqref{52} in the
form
\begin{align}\label{60}
\frac{d}{dt}\langle\hat a^{\dagger }\hat a\rangle &=-\kappa\langle
\hat a^{\dagger }\hat a\rangle + \varepsilon(\langle\hat
a^{\dagger 2}\rangle+\langle\hat a^2\rangle)\notag\\
&+\frac{\gamma_{c}}{1-4\varepsilon^2/\kappa^2}\langle\hat\sigma_{+}\hat
\sigma_{-}\rangle+\frac{4\gamma_{c}\varepsilon^2/\kappa^2}{(1-4\varepsilon^2/\kappa^2)^2}
\langle\hat\sigma_{z}\rangle.
\end{align}
On account of \eqref{30} and \eqref{32}, Eqs. \eqref{59} and
\eqref{60} reduce at steady state to
\begin{align}\label{61}
\langle\hat
a^2\rangle_{ss}&=\frac{2\varepsilon}{\kappa}\langle\hat
a^{\dagger}\hat
a\rangle_{ss}+\frac{\varepsilon}{\kappa}+\frac{\gamma_{c}\varepsilon/\kappa^2}
{1-4\varepsilon^2/\kappa^2}\notag\\
&-\frac{2\gamma_{c}\varepsilon/\kappa^2}
{(1+4\varepsilon^2/\kappa^2)(1-4\varepsilon^2/\kappa^2)}
\end{align}
and
\begin{equation}\label{62}
\langle\hat a^{\dagger}\hat
a\rangle_{ss}=\frac{\varepsilon}{\kappa}(\langle\hat
a^2\rangle_{ss}+\langle\hat a^{\dagger 2}\rangle_{ss}).
\end{equation}

Now with the aid of \eqref{61} and \eqref{62}, the mean photon
number of the cavity mode is found at steady state to be
\begin{align}\label{63}
\langle\hat a^{\dagger }\hat
a\rangle_{ss}&=\frac{2\varepsilon^2/\kappa^2}{1-4\varepsilon^2/\kappa^2}
-\frac{4\gamma_{c}\varepsilon^2/\kappa^3}{(1-4\varepsilon^2/\kappa^2)^2(1+4\varepsilon^2/\kappa^2)}\notag\\
&+\frac{(\gamma_{c}/2\kappa)(4\varepsilon^2/\kappa^2)}{(1-4\varepsilon^2/\kappa^2)^2}.
\end{align}
We observe that the first term in Eq. \eqref{63} represents the
mean photon number of the signal light in the absence of the
two-level atom ($\gamma_{c}=0$), the second term corresponds to
the mean number of absorbed signal photons, and the last term
represents the mean number of photons emitted by the two-level
atom. Therefore, the cavity mode is a superposition of the signal
light with a mean photon number
\begin{equation}\label{64}
\bar
n_{s}=\frac{2\varepsilon^2/\kappa^2}{1-4\varepsilon^2/\kappa^2}
-\frac{4\gamma_{c}\varepsilon^2/\kappa^3}{(1-4\varepsilon^2/\kappa^2)^2(1+4\varepsilon^2/\kappa^2)}
\end{equation}
and the fluorescent light with a mean photon number
\begin{equation}\label{65}
\bar
n_{f}=\frac{2\gamma_{c}\varepsilon^2/\kappa^3}{(1-4\varepsilon^2/\kappa^2)^2}.
\end{equation}
Expression \eqref{64} indicates that the presence of the two-level
 atom leads to a decreases in the mean photon number of the
signal light. Upon adding the last two terms in \eqref{63}, the
mean photon number of the cavity mode takes the form
\begin{equation}\label{66}
\langle\hat a^{\dagger }\hat
a\rangle_{ss}=\frac{2\varepsilon^2/\kappa^2}{1-4\varepsilon^2/\kappa^2}
-\frac{2\gamma_{c}\varepsilon^2/\kappa^3}
{(1-4\varepsilon^2/\kappa^2)(1+4\varepsilon^2/\kappa^2)}.
\end{equation}
Since the second term is negative, we conclude that the mean
number of photons absorbed by the two-level atom is greater than
the mean number of emitted photons.

Applying Eq. \eqref{62} in \eqref{50}, we get
\begin{equation}\label{67}
\Delta \hat a_{\pm}^2=1\pm \frac{\kappa}{\varepsilon}(1\pm
\frac{2\varepsilon}{\kappa})\langle\hat a^{\dagger}\hat
a\rangle_{ss}.
\end{equation}
Now introducing \eqref{66} into \eqref{67}, the quadrature
variance for the cavity mode is found to be
\begin{equation}\label{68}
\Delta\hat
a_{+}^2=1+\frac{(2\varepsilon/\kappa)(1-\gamma_{c}/\kappa)+8\varepsilon^3/\kappa^3}
{(1-2\varepsilon/\kappa)(1+4\varepsilon^2/\kappa^2)}
\end{equation}
and
\begin{equation}\label{69}
\Delta\hat
a_{-}^2=1-\frac{(2\varepsilon/\kappa)(1-\gamma_{c}/\kappa)+8\varepsilon^3/\kappa^3}
{(1+2\varepsilon/\kappa)(1+4\varepsilon^2/\kappa^2)}.
\end{equation}
We recall that in the bad-cavity limit, the cavity damping
constant $\kappa$ is much greater than the cavity atomic decay
rate $\gamma_{c}$, i.e., $\gamma_{c}/\kappa\ll 1$. In view of
this, we see that $1-\gamma_{c}/\kappa$ is positive. Moreover, we
note that Eq. (9) has a well-behaved solution provided that
$2\varepsilon/\kappa<1$. This implies that $1-2\varepsilon/\kappa$
is positive. Now on account of the fact that $1-\gamma_{c}/\kappa$
and $1-2\varepsilon/\kappa$ are positive, we see that $\Delta
a_{+}^2>1$ and $\Delta a_{-}^2<1$. Therefore the cavity mode is in
a squeezed state and the squeezing occurs in the minus quadrature.
In Fig. 5, we plot Eq. (69) versus $\varepsilon/\kappa$. This plot
also shows that the cavity mode is in a squeezed state and the
degree of squeezing increases with $\varepsilon/\kappa$.
\begin{figure}
\includegraphics [height=6cm,angle=0]{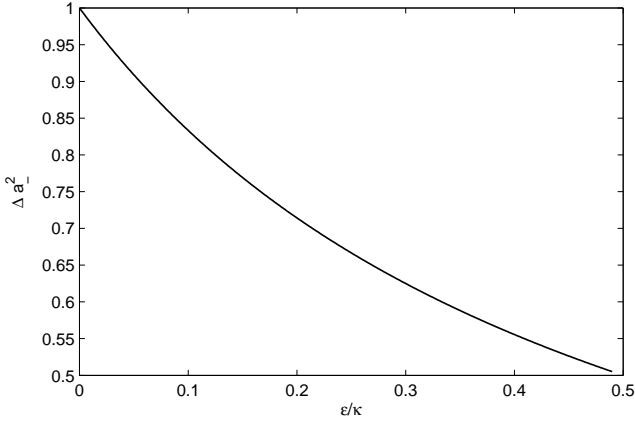}
\caption{Plots of the quadrature variance of the cavity mode [Eq.
\eqref{69}] versus $\varepsilon/\kappa$ for
$\gamma_{c}/\kappa=0.01$.}
\end{figure}
\begin{figure}
\includegraphics [height=6cm,angle=0]{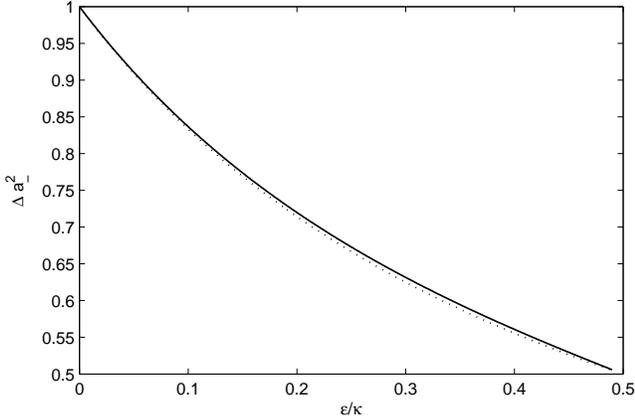}
\caption{Plots of the quadrature variance of the signal light [Eq.
\eqref{71}] versus $\varepsilon/\kappa$ in the presence of the
two-level atom with $\gamma_{c}/\kappa=0.01$ (solid curve) and in
the absence of the two-level atom, i.e, for $\gamma_{c}=0$ (dotted
curve).}
\end{figure}
On the other hand, using \eqref{67} and \eqref{64}, we find the
quadrature variance of the signal light to be of the form
\begin{equation}\label{70}
\Delta\hat
a_{+}^2=1+\frac{(2\varepsilon/\kappa)(1-16\varepsilon^4/\kappa^4-2\gamma_{c}/\kappa)}
{(1-2\varepsilon/\kappa)(1-16\varepsilon^4/\kappa^4)}
\end{equation}
and
\begin{equation}\label{71}
\Delta\hat
a_{-}^2=1-\frac{(2\varepsilon/\kappa)(1-16\varepsilon^4/\kappa^4-2\gamma_{c}/\kappa)}
{(1+2\varepsilon/\kappa)(1-16\varepsilon^4/\kappa^4)}.
\end{equation}
We note that
$16\varepsilon^4/\kappa^4=(2\varepsilon/\kappa)^4<2\varepsilon/\kappa<1$
and with $\gamma_{c}/\kappa$ being of the order of 0.01, we assert
that $1-16\varepsilon^4/\kappa^4-2\gamma_{c}/\kappa$ is positive.
We then see that for the signal light $\Delta a_{+}^2>1$ and
$\Delta a_{-}^2<1$ and hence the squeezing occurs in the minus
quadrature. In Fig. 6, we plot Eq. (71) versus
$\varepsilon/\kappa$ in the presence and in the absence of the
two-level atom. We see from this figure that the degree of
squeezing of the signal light slightly decreases due to the
presence of the two-level atom. We also see that the degree of
squeezing increases as $\varepsilon/\kappa$ increases. It is well
known that the signal light consists of highly correlated pairs of
photons and this correlation is responsible for the squeezing of
this light. Since the two-level atom absorbs a single photon at a
time, it somewhat destroys the correlations between signal photon
pairs. This leads to the decrease in the degree of squeezing of
the signal light.

It is also interesting to check if the fluorescent light emitted
by the two-level atom is in a squeezed state. To this end,
applying Eqs. \eqref {67} and \eqref{65} the quadrature variance
of the fluorescent light can be expressed as
\begin{equation}\label{72}
\Delta\hat a_{+}^2=1+\frac{2\gamma_{c}\varepsilon/\kappa^2}
{(1-2\varepsilon/\kappa)(1-4\varepsilon^2/\kappa^2)}
\end{equation}
and
\begin{equation}\label{73}
\Delta\hat a_{-}^2=1-\frac{2\gamma_{c}\varepsilon/\kappa^2}
{(1+2\varepsilon/\kappa)(1-4\varepsilon^2/\kappa^2)}.
\end{equation}
\begin{figure}
\includegraphics [height=6cm,angle=0]{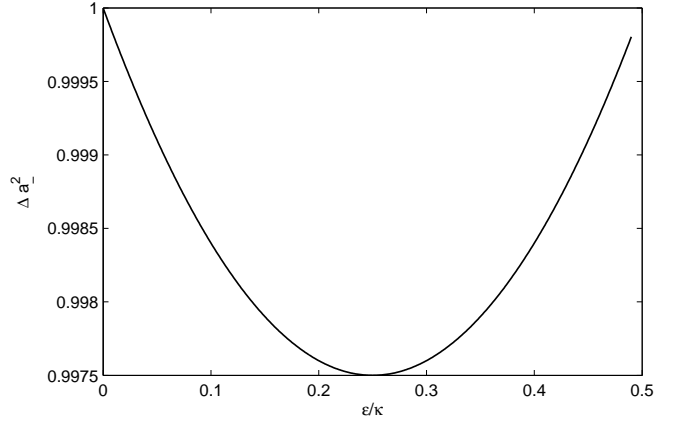}
\caption{Plots of the quadrature variance of the fluorescent light
[Eq. \eqref{73}] versus $\varepsilon/\kappa$ for
$\gamma_{c}/\kappa=0.01$.}
\end{figure}
We note from this result that the fluorescent light is in a
squeezed state. Fig. 7 indicates that the degree of squeezing of
the fluorescent light is very small.

\section{power spectrum of the cavity mode}
We finally determine the power spectrum of the cavity mode. The
power spectrum of the cavity mode can be expressed as
\begin{equation}\label{74}
S(\omega)=2Re\int_{0}^{\infty}\langle\hat a^{\dagger}(t)\hat
a(t+\tau)\rangle_{ss}e^{i\omega\tau}d\tau.
\end{equation}
With the aid of Eqs. \eqref{6b} and \eqref{7a}, we easily get
\begin{equation}\label{75}
\frac{d}{dt}\langle\hat a\rangle=-\frac{\kappa}{2}\langle\hat
a\rangle+\varepsilon \langle\hat
a^{\dagger}\rangle-g\langle\hat\sigma_{-}\rangle.
\end{equation}
Applying \eqref{75} and its complex conjugate, one can write
\begin{equation}\label{76}
\frac{d}{dt}\alpha_{\pm}=-\mu_{\mp}\alpha_{\pm}-gz_{\pm},
\end{equation}
in which
$\mu_{\mp}=\kappa(\frac{1}{2}\mp\frac{\varepsilon}{\kappa})$ and
$\alpha_{\pm}=\langle\hat a\rangle\pm \langle\hat
a^{\dagger}\rangle$. A formal solution of this equation can be
written as
\begin{equation}\label{77}
\alpha_{\pm}(t+\tau)=\alpha_{\pm}(t)e^{-\mu_{\mp}\tau}-g~e^{-\mu_{\mp}\tau}
\int_{0}^{\tau}
e^{\mu_{\mp}\tau^{\prime}}z_{\pm}(t+\tau^{\prime})d\tau^{\prime},
\end{equation}
so that on account of Eq. \eqref{27}, we have
\begin{align}\label{78}
\alpha_{\pm}(t+\tau)&=\alpha_{\pm}(t)e^{-\mu_{\mp}\tau}-g~z_{\pm}(t)e^{-\mu_{\mp}\tau}\notag\\
&\times \int_{0}^{\tau}
e^{-(\lambda_{\pm}-\mu_{\mp})\tau^{\prime}}d\tau^{\prime}.
\end{align}
Upon performing the integration, we obtain
\begin{equation}\label{79}
\alpha_{\pm}(t+\tau)=\alpha_{\pm}(t)e^{-\mu_{\mp}\tau}+\frac{g~z_{\pm}(t)}
{\lambda_{\pm}-\mu_{\mp}}
(e^{-\lambda_{\pm}\tau}-e^{-\mu_{\mp}\tau}).
\end{equation}
It then follows that
\begin{align}\label{80}
\langle\hat a(t+\tau)\rangle &=\frac{1}{2}(\langle\hat
a\rangle+\langle\hat
a^{\dagger}\rangle)e^{-\mu_{-}\tau}+\frac{1}{2}(\langle\hat
a\rangle-\langle\hat a^{\dagger}\rangle)e^{-\mu_{+}\tau}\notag\\
&+g\frac{\langle\hat\sigma_{-}\rangle+\langle\hat\sigma_{+}\rangle}{2(\lambda_{+}-\mu_{-})}
(e^{-\lambda_{+}\tau}-e^{-\mu_{-}\tau})\notag\\
&+g\frac{\langle\hat\sigma_{-}\rangle-\langle\hat\sigma_{+}\rangle}{2(\lambda_{-}-\mu_{+})}
(e^{-\lambda_{-}\tau}-e^{-\mu_{+}\tau}).
\end{align}
and application of the quantum regression theorem leads to
\begin{align}\label{81}
\langle\hat a^{\dagger}(t)\hat
a(t+\tau)\rangle_{ss}&=N_{1}e^{-\mu_{-}\tau}+N_{2}e^{-\mu_{+}\tau}\notag\\
&+N_{3}e^{-\lambda_{+}\tau} +N_{4}e^{-\lambda_{-}\tau},
\end{align}
where
\begin{figure}
\includegraphics [height=6cm,angle=0]{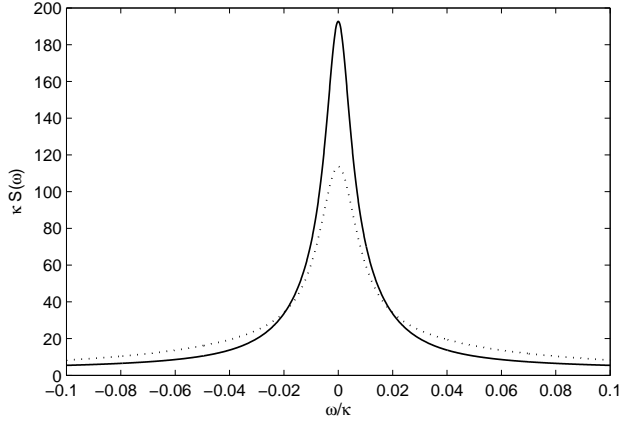}
\caption{Plots of the power spectrum of the cavity mode [Eq.
\eqref{83}] versus $\omega/\kappa$ for $\gamma_{c}/\kappa=0.01$,
for $\varepsilon/\kappa=0.25$ (solid curve) and for
$\varepsilon/\kappa=0.35$ (dotted curve).}
\end{figure}
\begin{figure}
\includegraphics [height=6cm,angle=0]{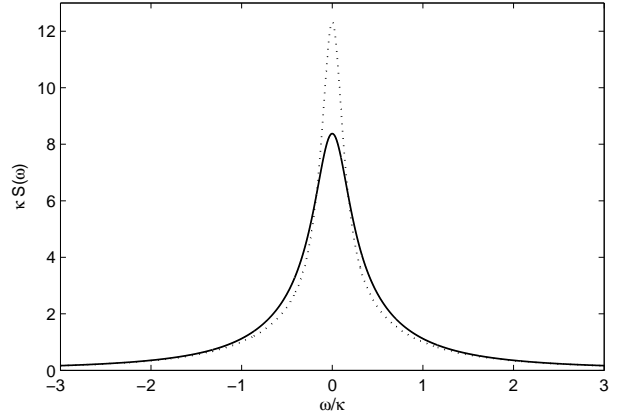}
\caption{Plot of the power spectrum of the signal light [Eq.
\eqref{84}] versus $\omega/\kappa$ for $\varepsilon/\kappa=0.25$
(solid curve) and for $\varepsilon/\kappa=0.35$ (dotted curve).}
\end{figure}
\begin{align}\label{82}
N_{1}=\frac{1}{2}(\langle\hat a^{\dagger}\hat
a\rangle_{ss}+\langle\hat a^{\dagger
2}\rangle_{ss}+\frac{g}{2}\frac{\langle\hat
a^{\dagger}\hat\sigma_{-}\rangle_{ss}+\langle\hat
a^{\dagger}\hat\sigma_{+}\rangle_{ss}}{\lambda_{+}-\mu_{-}})\notag\\
N_{2}=\frac{1}{2}(\langle\hat a^{\dagger}\hat
a\rangle_{ss}-\langle\hat a^{\dagger
2}\rangle_{ss}+\frac{g}{2}\frac{\langle\hat
a^{\dagger}\hat\sigma_{-}\rangle_{ss}-\langle\hat
a^{\dagger}\hat\sigma_{+}\rangle_{ss}}{\lambda_{-}-\mu_{+}})\notag\\
N_{3}=\frac{g}{2}\frac{\langle\hat
a^{\dagger}\hat\sigma_{-}\rangle_{ss}+\langle\hat
a^{\dagger}\hat\sigma_{+}\rangle_{ss}}{\lambda_{+}-\mu_{-}}
\notag\\
N_{4}=\frac{g}{2}\frac{\langle\hat
a^{\dagger}\hat\sigma_{-}\rangle_{ss}-\langle\hat
a^{\dagger}\hat\sigma_{+}\rangle_{ss}}{\lambda_{-}-\mu_{+}}.
\end{align}
On account of \eqref{81}, the normalized power spectrum of the
cavity mode turns out to be
\begin{align}\label{83}
S(\omega)&=\frac{\kappa(\frac{1}{2}+\frac{\varepsilon}{\kappa})/4\pi}
{\kappa^2(\frac{1}{2}+\frac{\varepsilon}{\kappa})^2+\omega^2}
+\frac{\kappa(\frac{1}{2}-\frac{\varepsilon}{\kappa})/4\pi}
{\kappa^2(\frac{1}{2}-\frac{\varepsilon}{\kappa})^2+\omega^2}\notag\\
&+\frac{\Gamma(\frac{1}{2}+\frac{\varepsilon}{\kappa})/4\pi}
{\Gamma^2(\frac{1}{2}+\frac{\varepsilon}{\kappa})^2+\omega^2}
+\frac{\Gamma(\frac{1}{2}-\frac{\varepsilon}{\kappa})/4\pi}
{\Gamma^2(\frac{1}{2}-\frac{\varepsilon}{\kappa})^2+\omega^2}.
\end{align}
We identify that
\begin{equation}\label{84}
S(\omega)=\frac{\kappa(\frac{1}{2}+\frac{\varepsilon}{\kappa})/2\pi}
{\kappa^2(\frac{1}{2}+\frac{\varepsilon}{\kappa})^2+\omega^2}
+\frac{\kappa(\frac{1}{2}-\frac{\varepsilon}{\kappa})/2\pi}
{\kappa^2(\frac{1}{2}-\frac{\varepsilon}{\kappa})^2+\omega^2}
\end{equation}
is the power spectrum of the signal light. The last two terms in
Eq. \eqref{83} represent the power spectrum of the fluorescent
light, which is the same as Eq. \eqref{35}. Since the expression
for the spectrum of the signal light does not contain
$\gamma_{c}$, the presence of the two-level atom does not affect
the width of this spectrum. In Fig. 8, we plot the power spectrum
of the cavity mode versus $\omega/\kappa$ for different values of
$\varepsilon/\kappa$. These plots show that the width of the power
spectrum increases as the degree of squeezing increases. When the
value of $\varepsilon/\kappa$ increases from $0.25$ to $0.35$, the
half width increases from $0.0072$ to $0.0108$. In addition, in
Fig. 9 we plot the power spectrum of the signal light versus
$\omega/\kappa$ for different values of $\varepsilon/\kappa$.
These plots indicate that the width of the spectrum decreases  as
the degree of squeezing increases. The half width of the spectrum
decreases from 0.3168 to 0.1766 as $\varepsilon/\kappa$ increases
from 0.25 to 0.35.
\section{Conclusion}
We have studied a degenerate parametric oscillator with a
two-level atom applying the Heisenberg and quantum Langevin
equations in the bad-cavity limit. We have obtained the mean
photon number, the quadrature variance, and the power spectrum for
the cavity mode, for the signal light, and for the fluorescent
light. In addition, we have determined the second-order
correlation function for the fluorescent light. The method we have
used enables us to investigate both the atomic fluorescence and
the quantum statistical properties of the cavity mode.

We have found that the photons in the fluorescent light are
antibunched. Unlike the power spectrum of the fluorescent light
from a two-level atom driven by a strong coherent light, the power
spectrum of the fluorescent light in this case turns out to be a
single peak. It is found that the width of the spectrum increases
with $\varepsilon/\kappa$. Moreover, we have seen that the
fluorescent light is in a squeezed state with a very small amount
of squeezing.

On the other hand, the presence of the two-level atom leads to a
decrease in the mean photon number and in the degree of squeezing
of the signal light. However, the presence of the two-level atom
has no effect on the spectrum of the signal light.

\begin{acknowledgments}
One of the authors (Eyob Alebachew) is grateful to the Abdus Salam
ICTP for the financial support under the affiliated center (AC-14)
at the Department of Physics of the Addis Ababa University.
\end{acknowledgments}

\end{document}